\documentclass[a4paper,12pt]{article}
\usepackage[utf8]{inputenc}
\usepackage{amsmath}
\usepackage{amssymb}
\usepackage{amsfonts}
\usepackage{amssymb,amsmath}
\usepackage{setspace}
\usepackage{fullpage}
\usepackage[pdftex]{graphicx}  
\usepackage{float}
\usepackage{subfigure}
\usepackage{hyperref}
\usepackage{cite}
\usepackage{color}
\usepackage{verbatim}
\usepackage[font={small,it}]{caption}
\def\beq{\begin{equation}}
\def\eeq{\end{equation}}
\def\bea{\begin{eqnarray}}
\def\eea{\end{eqnarray}}

\begin{document}

\bibliographystyle{OurBibTeX}

\begin{titlepage}

\vspace*{-25mm}
\begin{flushright}
ADP-13-05/T825 \\ 
\end{flushright}

\begin{center}
{ \sffamily \Large \bf
Fine Tuning in the
Constrained Exceptional Supersymmetric Standard Model
}
\\[8mm]
P.~Athron$^{a}$,
\footnote{E-mail: \texttt{peter.athron@adelaide.edu.au}}
Maien~Binjonaid$^{b,c}$,
\footnote{E-mail: \texttt{mymb1a09@soton.ac.uk, maien@ksu.edu.sa}}
and S.F.~King$^{b}$
\footnote{E-mail: \texttt{king@soton.ac.uk}}
\\[3mm]
{\small\it
$^a$ ARC Centre of Excellence for Particle Physics at the Terascale, \\
School of Chemistry and Physics,
The University of Adelaide, \\
Adelaide, SA 5005, Australia. \\[2mm]
$^b$ School of Physics and Astronomy, University of Southampton,\\
Southampton, SO17 1BJ, U.K.\\[2mm]
$^c$ Department of Physics and Astronomy, King Saud University,\\
Riyadh 11451, P. O. Box 2455, Saudi Arabia
}\\[1mm]
\end{center}
\vspace*{0.5cm}

\begin{abstract}
\noindent
Supersymmetric unified models in which the $Z^\prime$
couples to the Higgs doublets, as in the $E_6$ class of models,
have large fine tuning dominated by the experimental mass limit on the $Z^\prime$.
To illustrate this we investigate the degree of fine tuning throughout the parameter
space of the Constrained Exceptional Supersymmetric Standard Model (cE$_6$SSM)
that is consistent with a Higgs mass $m_h \sim 125$ GeV.
Fixing $\tan\beta =10$, 
and taking specific values of the mass of the $Z^\prime$ boson, with $M_{Z'}\sim 2-4$ TeV.
We find that the minimum fine tuning is set predominantly from the mass of $Z^{\prime}$ and varies from $\sim 200-400$ as 
we vary $M_{Z'}$ from $\sim~ 2-4$ TeV.
However, this is significantly lower than the fine tuning in the Constrained Minimal Supersymmetric Standard Model (cMSSM),
of $\mathcal{O}$(1000), arising from the large stop masses required to achieve the Higgs mass.

\end{abstract}

\end{titlepage}
\newpage


\section{Introduction}

The Large Hadron Collider (LHC) has been accumulating data since 2009
with no observation of new physics beyond the standard model (BSM) so
far, placing strong limits on new coloured states in
extensions of the standard model. For example, in supersymmetric
(SUSY) models there are strong experimental limits on the first and second
generation squark and gluino masses \cite{:2012uu, Aad:2012hm} which imply that
they must be at least an order of magnitude larger than
the electroweak (EW) scale. Within
constrained versions of SUSY, where the stop masses are linked
to first and second generation squarks masses, this can considerably
increase fine tuning since the EW scale is very sensitive to stop
masses, through the electroweak symmetry breaking conditions.

At the same time Atlas and CMS have recently observed a new state
consistent with a Standard-Model-like Higgs boson at $m_h = 125-126$
GeV \cite{:2012gk,:2012gu}, which is within the range for it to be
consistent with the lightest Higgs in supersymmetric models.  
In the minimal supersymmetric standard model (MSSM) this
introduces further tension with naturalness since the light Higgs mass
at tree-level is bounded from above by the $Z$ boson mass ($M_Z$).
The large radiative contributions from stops needed to raise it to the
observed value typically imply very large fine tuning. 
For example the constrained MSSM (cMSSM) \cite{Chung:2003fi} has been shown to 
require fine tuning of $\mathcal{O}(1000)$ if it is to
contain a 125 GeV Higgs mass \cite{Cassel:2011zd,
  Ghilencea:2012gz}. 
  
Here we consider fine tuning in an alternative class of constrained SUSY models which 
involves both an extra singlet field, denoted $S$, and an extra 
$U(1)$ gauge symmetry at low energy (TeV scale). As the singlet acquires a VEV, denoted $s$, it produces a $\mu$ term, denoted $ \mu_{\text{eff}}$, and it breaks the extra $U(1)$ gauge symmetry, giving rise to a massive $Z^{\prime}$ boson.
Such models can increase the tree-level physical Higgs boson mass above the $M_Z$
limit of the MSSM, due to both F-term contributions of the singlet and the D-term contributions
associated with the $Z'$, allowing lighter stop masses and hence reducing fine tuning due to stop loops.
The exceptional supersymmetric standard model (E$_6$SSM) \cite{King:2005jy, King:2005my}
is an example of such a model, inspired by the $E_6$ group. At tree-level, the 
light Higgs mass is given as,
\begin{equation} \label{mh}
m_h^2 \approx \underbrace{ \underbrace{ \underbrace{M_Z^2 \cos^2 2\beta}_{\text{MSSM}} +  \frac{ \lambda^2}{2} v^2 \sin^2 2 \beta }_{\text{NMSSM}} +  \frac{M_Z^2}{4}(1 + \frac{1}{4} \cos 2 \beta)^2 }_{\text{E$_6$SSM}} + \Delta m_h^2,
\end{equation}
where, $\tan\beta = \frac{v_2}{v_1}$ is the ratio between the two Higgs doublets' vacuum expectation values (VEVs), $\lambda$ is the Yukawa coupling of the singlet field to the Higgs doublets, and $\Delta m_h^2$ represents loop corrections.

Indeed, Eq.~\ref{mh} shows that the E$_6$SSM allows larger tree-level Higgs
masses than the NMSSM \cite{genNMSSM2}, which in turn allows larger tree-level Higgs
masses than the MSSM.  This means that the E$_6$SSM does not rely on
such a large a contribution from the radiative correction term $\Delta
m_h^2$ in order to reproduce the Higgs mass. As a result
the E$_6$SSM permits lower stop masses than either the NMSSM or the
MSSM. In addition the $\lambda$ coupling in the E$_6$SSM
  can be larger at low energies, while still remaining perturbative
  all the way up to the GUT scale, than is the case in the NMSSM.

One might conclude that this should lead to lower fine tuning in
the E$_6$SSM than either the NMSSM or MSSM, since the large stop
masses are usually the main source of fine tuning in SUSY
models. However, the origin of the extra term in Eq.~\ref{mh} is due to
D-terms arising from the coupling of the Higgs doublets to the extra
$U(1)$ gauge symmetry, and such D-terms also contribute to the
minimisation conditions of the Higgs doublets. Indeed, as we shall
discuss, one of the minimisation conditions of the E$_6$SSM can be
written in the form, \beq
\label{mz0}
c \frac{M_Z^2}{2}=-\mu_{\text{eff}}^2+  \frac{ (m_d^2 - m_u^2 \tan^2\beta) }{ \tan^2\beta - 1 } +d \frac{M_{Z'}^2}{2},
\eeq
where $c,d$ are functions of $\tan \beta$ which are of order $\sim \mathcal{O}(1)$,
$m_d^2, m_u^2$ are soft Higgs mass squared parameters
and $ \mu_{\text{eff}}$ arises from the singlet VEV.
Written in this form it is clear that the D-terms are a double edged sword since they also
introduce a new source of tree-level fine tuning, due to the $Z'$ mass squared term in Eq.~\ref{mz0}, which 
will increase quadratically as $M_{Z'}^2$, eventually coming to dominate the fine tuning for large enough 
values of $M_{Z'}$.
This tree-level fine tuning can be compared to that due to $\mu_{\text{eff}}$ which typically requires this
parameter to be not much more than 200 GeV, and similar limits also apply to $M_{Z'}$.
With the current CMS experimental mass limit for the $Z'$ in the E$_6$SSM of 
$M_{Z'}\gtrsim 2.08$ TeV \cite{Chatrchyan:2012it} it is clear that there is 
already a significant, perhaps dominant, amount of fine tuning due to the $Z'$ mass limit.

In this paper we investigate this new and important source of fine tuning,
namely that due to the $M_{Z'}$ limit, and compare it to the usual other sources of fine tuning in the framework of
 the Constrained E$_6$SSM (cE$_6$SSM) \cite{Athron:2011wu,
  Athron:2010zz, Athron:2009bs, Athron:2009ue}.
 Although the impact of a SM-like
Higgs with $m_h \sim 125$ GeV on the parameters has recently been considered in
\cite{Athron:2012sq,Athron:2012pw}, fine tuning was not considered. 
In fact the present study here is the first time that fine tuning has been considered 
in any supersymmetric $E_6$ model with a low energy $Z'$.
To obtain the required Higgs mass in the cE$_6$SSM, it turns out that the SM
singlet field, $S$, must have a VEV $s \geq 5$ TeV as pointed
out in \cite{Athron:2012sq}. This corresponds to a mass of the $Z'$
boson predicted by the model of 1.9 TeV, which almost reaches the
experimental bound of 2 TeV \cite{Chatrchyan:2012it}. Thus, all the
parameter space we study respects the experimental limit on
$M_{Z'}.$ Fixing $\tan\beta =10$, 
and taking specific values of the mass of the $Z^\prime$ boson, $M_{Z'}$, ranging from 1.9 to 3.8
TeV we find that the current minimum fine tuning in the cE$_6$SSM,
consistent with a Higgs mass $m_h \sim 125$ GeV, 
varies from $\sim 200 - 400$, 
and is already dominated by the $M_{Z'}$ limit. 
However, this is significantly lower than the fine tuning in the cMSSM
of $\mathcal{O}(1000)$ arising from the large stop masses required to achieve the Higgs mass.

The rest of the paper is organised as follows: Section two provides a
short overview of the E$_6$SSM.  Then, the scalar Higgs potential and
the electroweak symmetry breaking (EWSB) conditions are discussed in
Section three. In Section four we discuss the fine tuning measure we
use, and derive a fine tuning master formula for the E$_6$SSM with a
brief description of our semi-numerical procedure of calculating fine
tuning.  Section five is where we present our results and discussion,
then we conclude the study in Section six.

\section{The E$_6$SSM}

The Exceptional Supersymmetric Standard Model (E$_6$SSM) is a
non-minimal supersymmetric extension of the SM, which provides a low
energy alternative to the MSSM and NMSSM. It is well
  motivated both from more fundamental theories due to its connection
  to $E_6$ GUTs, heterotic and F- string theory \cite{Callaghan:2012rv} and at the same time as a
  low energy effective model, providing solutions to phenomenological
  problems.  For instance, as mentioned in the Introduction, the E$_6$SSM allows
  a larger Higgs mass at tree-level than in both the MSSM and the
  NMSSM, thereby requiring smaller contributions from loops. In
addition it also solves the $\mu$ problem associated with the MSSM by
dynamically producing the $\mu$-term at the TeV scale, without
introducing the domain walls or tadpole problems that can appear in
the NMSSM.

The E$_6$SSM is based on the Exceptional Lie group $E_6$. This contains both of $SO(10)$ and $SU(5)$ as subgroups, 
\begin{align}
E_6 \rightarrow SO(10) &\times U(1)_{\psi}  \\
SO(10) \rightarrow SU(5) &\times U(1)_{\chi},
\end{align}
and hence also contains the Standard Model gauge group,
  which is a subgroup $SU(5)$. A linear combination of the two extra
$U(1)_{\psi}$ and $U(1)_{\chi}$ groups can survive to low energies,
where it is spontaneously broken by a SM singlet field, $S$. This generates the mass of the associated $Z^{\prime}$ boson and the
exotic quarks, as well as dynamically producing a $\mu_{eff}$
term. The model allows right-handed (RH) neutrinos to have Majorana
masses at some scale between the GUT and low scales. This is achieved
by choosing this linear combination to be,
\begin{align}
U(1)_N &= \frac{\sqrt{15}}{4}  U(1)_{\psi} + \frac{1}{4} U(1)_{\chi}
\end{align}
such that the RH neutrinos are not charged under $U(1)_N$, hence it is possible to explain the tiny neutrino masses via seesaw mechanisms. 

At low energies, the group structure of the model is that of the SM, along with the additional $U(1)_N$ symmetry,

\begin{align}
E_6 &\rightarrow SU(5) \times U(1)_N \\
SU(5) &\rightarrow SU(3)_c \times SU(2)_w \times U(1)_Y 
\end{align}
The matter content of the model is contained in the complete 27-dimensional representation which decomposes under $SU(5) \times U(1)_N$ to,

\begin{equation}
27_i \longrightarrow (10, 1)_i + (5^*, 2)_i + (5^*, -3)_i + (5, -2)_i + (1, 5)_i + (1, 0)_i
\end{equation}
Ordinary Quarks and Leptons are contained in the representations: $(10,1)$ and $(5^*,2).$ The Higgs doublets and exotic quarks are contained in $(5^*, -3)$ and $(5,-2).$ 
The singlets are contained in $(1,5)$, and finally the right handed neutrinos are included in $(1,0).$

Moreover, the model requires three 27 representations, hence $i =
1,2,3$, in order to ensure anomaly cancellation.  This means that
there are three copies of each field present in the model. However,
only the third generation (by choice) of the two Higgs doublets, and
the SM singlet acquire VEVs. The other two generations are called
inert. Furthermore, in order to keep gauge coupling unification,
non-Higgs fields that come from extra incomplete $27', \bar{27}'$
representations are added to the model. As a result, a $\mu^{\prime}$
term, which is not necessary related to the weak scale, is present in
the model.

The full superpotential consistent with the low energy gauge structure
of the E$_6$SSM contains includes both $E_6$ invariant invariant terms
and $E_6$ breaking terms, full details of which are given in
\cite{King:2005jy}.  However as in the MSSM it is necessary to forbid
proton decay and therefore a generalsation of R-parity should be
imposed, and additionally because the E$_6$SSM includes three
generations of every chiral superfield, there needs to be a
suppression of new terms which can induce flavour changing neutral
currents.  To achieve this we impose either a $Z_2^L$
symmetry\footnote{All superfields except the leptons and survival
  Higgs are even.} (Model I) or a $Z_2^B$ symmetry\footnote{All the
  exotic quark, lepton and survival Higgs superfields are odd while
  all the other superfields remain even.} (Model II) along with an
approximate $Z_2^H$ symmetry, under which all fields are odd except
for the third generation Higgs superfields, which may arise from a
family symmetry\cite{Howl:2008xz,Howl:2009ds}.

 The $Z_2^H$ invariant superpotential then reads,
\begin{eqnarray}
W_{\rm E_6SSM} &\approx &
\lambda_i \hat{S}(\hat{H}^d_{i}\hat{H}^u_{i})+\kappa_i \hat{S}(\hat{D}_i\hat{\overline{D}}_i)+
f_{\alpha\beta}\hat{S}_{\alpha}(\hat{H}_d \hat{H}^u_{\beta})+ 
\tilde{f}_{\alpha\beta}\hat{S}_{\alpha}(\hat{H}^d_{\beta}\hat{H}_u) \nonumber\\[2mm]
&&+\dfrac{1}{2}M_{ij}\hat{N}^c_i\hat{N}^c_j+\mu'(\hat{H}'\hat{\overline{H'}})+
h^{E}_{4j}(\hat{H}_d \hat{H}')\hat{e}^c_j+h_{4j}^N (\hat{H}_{u} \hat{H}')\hat{N}_j^c 
\nonumber \\[2mm]
&& + W_{\rm{MSSM}}(\mu=0),
\label{Eq:SupPot}
\end{eqnarray}

\noindent where the indices $\alpha, \beta = 1,2$ and $i = 1,2,3$ denote the generations. $S$ is the SM singlet field, $H_u,$ and $H_d$ are the Higgs doublet fields 
corresponding to the up and down types.  Exotic quarks and the additional non-Higgs fields are denoted by $D$ and $H'$ respectively. 

Finally to ensure that only third generation Higgs like fields get
VEVs a certain hierarchy between the Yukawa couplings must
exist. Defining $\lambda \equiv \lambda_3$, we impose
$\kappa_i,\lambda_i \gg
f_{\alpha\beta},\,\tilde{f}_{\alpha\beta},\,h^{E}_{4j},\,h_{4j}^N$. Moreover, we do not impose any unification of
the Yukawa couplings at the GUT scale.

\section{The Higgs potential and the EWSB conditions}

The scalar Higgs potential is,
\begin{equation}
\begin{split}
V(H_d, H_u, S) = & \ { \lambda^2 |S|^2 (|H_d|^2 + |H_u|^2 ) + \lambda^2 | H_d . Hu |^2 }\\
& + { \frac{ g_2^2 }{ 8 } (H^\dagger_d \sigma_a H_d + H^{\dagger}_u \sigma_a H_u ) (H^ {\dagger}_d \sigma_a H_d + H^{\dagger}_u \sigma_a H_u ) }\\
& + { \frac{ {g'}^2 }{ 8 } (|H_d|^2 - |H_u|^2 )^2 + \frac{ {g'}_1^2 }{ 2 } (Q_1 |H_d|^2 + Q_2 |H_u|^2 + Q_s |S|^2)^2 }\\
& + { m_s^2 |S|^2 + m_d^2 |H_d|^2 + m_u^2 |H_u|^2 }\\
& + { [\lambda A_{\lambda} S H_d . H_u + c.c. ] + \Delta_{\text{Loops}} }
\end{split}
\end{equation}
where, $g_2$, $g' (= \sqrt{3/5} g_1)$, and $g_1^\prime$ are the gauge couplings of $SU(2)_L, U(1)_Y$ (GUT normalized), 
and the additional $U(1)_N$, respectively. $Q_1=-3/\sqrt{40}, Q_2=-2/\sqrt{40},$ and $Q_s=5/\sqrt{40}$ are effective $U(1)_N$ charges of $H_u, H_d$ and $S$, respectively. 
$m_s$ is the mass of the singlet field, and $m_{u,d} \equiv m_{H_{u,d}}$.

The Higgs field and the SM singlet acquire VEVs at the physical minimum of this potential,

\begin{equation}
\langle H_d \rangle = \frac{1}{\sqrt{2}}   \begin{pmatrix} v_1 \\ 0 \end{pmatrix}, \ \ \  \langle H_u \rangle = \frac{1}{\sqrt{2}}   \begin{pmatrix} 0 \\ v_2 \end{pmatrix},  \langle S \rangle = \frac{s}{\sqrt{2}} , 
\end{equation}
\
\\
It is reasonable exploit the fact that $s \gg v$, which will help in simplifying our master formula 
for fine tuning as will be seen in Section 4. Then, from the minimisation conditions, 

\begin{equation}
 \frac{ {\partial V_{E_6SSM} } }{ \partial v_1 } = \frac{ \partial V_{E_6SSM} }{ \partial v_2 } = \frac{ \partial V_{E_6SSM} }{ \partial s } = 0 ,
\end{equation}
the Electroweak Symmetry Breaking (EWSB) conditions are,

\begin{equation} \label{mz}
\frac{M_Z^2}{2} = -\frac{1}{2} \lambda^2 s^2 +  \frac{ (m_d^2 - m_u^2 \tan^2\beta) }{ \tan^2\beta - 1 } + \frac{ g_1'^2 }{2} \left(Q_1 v_1^2 + Q_2 v_2^2 + Q_s s^2\right) \frac{ (Q_1 - Q_2 \tan^2\beta) }{ \tan^2\beta - 1 } 
\end{equation}

\begin{equation} \label{sin2b}
\sin 2\beta \approx \frac{ \sqrt{2} \lambda A_{\lambda} s }{ m_d^2 + m_u^2 + \lambda^2 s^2 + \frac{ g_1'^2 }{2} Q_s s^2 ( Q_1 + Q_2 ) },
\end{equation}

\begin{equation} \label{ms}
m_s^2 \approx -\frac{1}{2}  g_1'^2 Q_s^2 s^2 = - \frac{1}{2} M_{Z'}^2,
\end{equation}
where $M_Z^2=\frac{1}{4}({g'}^2+g_2^2)(v_2^2+v_1^2)$ and 
$M_{Z'}^2 \approx  g_1'^2 Q_s^2s^2$.

Eq.~\ref{mz} can be written in the form,
\beq
\label{mz2}
c \frac{M_Z^2}{2}=-\mu_{\text{eff}}^2+  \frac{ (m_d^2 - m_u^2 \tan^2\beta) }{ \tan^2\beta - 1 } +d \frac{M_{Z'}^2}{2},
\eeq
where $c,d$ are functions of $\tan \beta$ which are of order $\sim \mathcal{O}(1)$
and we have written $ \mu_{\text{eff}} = \frac{ \lambda s}{\sqrt{2}}$.
Written in this form it is clear that fine tuning 
will increase as $M_{Z'}$ increases. Another source of fine tuning is the large $|\mu_{\text{eff}}|$ term as mentioned in the introduction 
since satisfying Eq.~\ref{mz2} will require this term
to compensate for any increase in either the second term (term 2: $\sim m_u^2, m_d^2$) or the last term (term 3: $\sim M_{Z^\prime}^2$).

The increasing experimental limits on $M_{Z^{\prime}} (\sim s)$ results in constraining 
the parameter space of the E$_6$SSM such that only relatively large values of $m_0$ and $m_{1/2}$ result 
in successful solutions to the EWSB conditions (Fig.~\ref{Fig:s5}-~\ref{Fig:s10}). 
\

Moreover, imposing universal boundary conditions, which is what characterises the cE$_6$SSM, means that all low energy SUSY parameters can be expanded in terms of a few
GUT-scale universal and fundamental input parameters, namely,

\begin{equation}
 m_0, \ \ m_{1/2}, \ \ A, \ \ \lambda_i(0), \ \ \kappa_i(0), \ \ h_{t,b,\tau}(0)
\end{equation}
where, $m_0,  m_{1/2}$ and $A$  are a universal scalar mass, a universal gaugino mass, and a universal trilinear coupling, respectively, and $(0)$ means
taking the parameter at the GUT scale (in the Results section, we refer to $\lambda_3(0)$ and $\kappa_{1,2,3}(0)$ as $\lambda_0$ and $\kappa_0$, respectively).

This is accomplished by using the one-loop RGEs of the scalar masses, so that one can express $m_{H_u}^2$ at the SUSY scale, $M_S$, as,

\begin{equation}
 m_{H_u}^2 (M_S) = z_1 m_0^2 + z_2 m_{1/2}^2 + z_3 A^2 + z_4 m_{1/2} A.
\end{equation}
Then, it is possible to write,

\begin{equation}
 \frac{M_Z^2}{2} \approx \sum_{i=1}^n F_i z_i a_i^2
\end{equation}
where, $a$ denotes the fundamental parameters, $z$ is the coefficient corresponding to each parameter, and is calculated numerically. $F$ is
some factor, possibly, involving $\tan \beta$. 

Whence, one can calculate (analytically or numerically) the sensitivity of $M_Z$ to each fundamental parameter, and this leads us to fine tuning.

\section{Fine tuning and the master formula}
To study the degree of fine tuning, a quantitative measure needs to be applied. Here we use the conventional fine tuning measure \cite{Ellis:1986yg, Barbieri:1987fn}, where the fractional 
change in the observable is calculated for a given fractional change in the input parameter,

\begin{equation} \label{delta}
  \Delta_a = \left| \frac{\partial \ln M_Z }{\partial \ln a} \right|,
\end{equation}
where $M_Z$ is the mass of the $Z$ boson\footnote{Note that some authors choose $M_Z^2$ instead of $M_Z$. Both measures can be easily linked since $\frac{1}{2}\Delta_a (M_Z^2) = \Delta_a (M_Z)$. Our choice was made to enable straightforward comparisons with the results in ~\cite{Ghilencea:2012gz}.} and $a$ is one of
the fundamental parameter in the set
$\{m_0,m_{1/2},A,\lambda(0),\kappa(0)\}.$ 

For example, $\Delta_a = 10$ and $200$ correspond to a $10\%$ and $0.5\%$
tuning in the parameter $a$, respectively. Moreover, for a given point
in the parameter space, fine tuning is the maximum value of fine
tuning in the set $\{ \Delta_a \}$, and is denoted $\Delta_{max}$ (or
simply $\Delta$).
 
This measure has been used extensively within the literature e.g. \cite{deCarlos:1993yy, deCarlos:1993ca,
  Chankowski:1997zh, Agashe:1997kn, Wright:1998mk, Kane:1998im,
  Bastero-Gil:1999gu, Feng:1999zg, Allanach:2000ii, Dermisek:2005ar,
  Barbieri:2005kf, Allanach:2006jc, Gripaios:2006nn, Dermisek:2006py,
  Barbieri:2006dq, Kobayashi:2006fh, Perelstein:2012qg,
  Antusch:2012gv,Cheng:2012pe, CahillRowley:2012rv, Ross:2012nr,
  Basak:2012bd, Kang:2012sy}.

\subsection{Alternative tuning measures}
Some concerns have been raised in the literature regarding the use of
this measure and its use is not universal with a number of
alternative measures having been introduced and applied \cite{Anderson:1994dz, Anderson:1994tr,
  Anderson:1995cp,Anderson:1996ew,Ciafaloni:1996zh,Chan:1997bi,
  Barbieri:1998uv,
  Giusti:1998gz,Casas:2003jx,Casas:2004uu,Casas:2004gh,Casas:2006bd,Kitano:2005wc,
  Athron:2007ry, Baer:2012up}. The $\{\Delta_a\}$ measure the
sensitivity of the parameters to the observable and as such are very
dependent on the parametrization chosen.  In particular whether one
takes $p_i$ to be the parameter or instead chooses $a = p_i^2$
introduces a factor two difference, and this factor two will then
appear for every point in the parameter space. To remove this {\it
  global sensitivity} one can choose some normalisation
\cite{Anderson:1994dz,
  Anderson:1994tr,Anderson:1995cp,Anderson:1996ew} on the $\Delta_ a$,
however this then introduces questions about the bounds on the
parameters and the probability is not clearly defined or understood.
     
 Additionally the overall tuning is chosen by taking $\Delta$ as
 the maximum of the individual sensitivities $\{\Delta_a\}$, but a
 proposed alternative is to combine them in quadrature, like
 uncorrelated
 errors\cite{Casas:2003jx,Casas:2004uu,Casas:2004gh,Casas:2006bd}.
 Clearly these measures can differ substantially, but it is not
 obvious which should be chosen. A new measure\cite{Athron:2007ry}
 defined tuning\footnote{This measure also allows one to combine
   several observables and had a normalised version of the tuning
   measure to deal with global sensitivity in a similar manner
   to\cite{Anderson:1994dz,
     Anderson:1994tr,Anderson:1995cp,Anderson:1996ew}, but with a
   slightly different normalisation and interpretation in terms of
   probabilities.} as the ratio of the parameter space volume (defined
 by fixed dimensionless variations in the parameters) to the same
 volume with the additional constraint that the dimensional variations
 of the observable are no greater than those of the parameters.  As
 such this measure automatically combined the tuning from each
 parameter into a single tuning defined in terms of parameter space
 volume.  For simple cases studied it was shown that this new measure
 was in greater agreement with the conventional measure than the
 alternative where the sensitivities are combined in quadrature, which
 might be understood as being due to large correlations between the
 individual sensitivities.
 
Finally all the measures described so far define tuning as a theoretical feature of a
point in parameter space, measuring how natural a point is. As such these measures quantify
how natural phenomenologically acceptable points are once experimental limits have ruled out
points which were initially favoured as being natural (or more natural).
Instead within Bayesian analyses natural expectations for parameter space points, 
given by the prior distribution, are combined with experimental data to determine 
the probability defined as a degree of belief. If one must fine tune the parameters 
to get the measured values of observables correct, then this will correspond to 
only a tiny fraction of the total integrated prior volume, and therefore 
fine tuned scenarios should be automatically penalised.  However in
practice in MSSM studies $M_Z$ is often fixed to it's experimental
value at the outset, reducing the dimensionality of the parameter
space and missing the fine tuning. To fix this one can start off with
a full set of parameters with the chosen prior distribution,
unconstrained by EWSB requirements and then perform a Jacobian
transformation\cite{Allanach:2007qk, Cabrera:2008tj, Ghilencea:2012gz,
  Ghilencea:2012qk, Fichet:2012sn}.  The Jacobian factor accounts for
the missed fine tuning and introduces similar derivatives as those
appearing the sensitivity criterion, so it then appears as an
effective ''fine tuning prior''.

In the MSSM the conventional measure of fine tuning is numerically
very close to this effective fine tuning prior (see e.g. \cite{
  Ghilencea:2012gz}) and has sometimes been used directly as a fine
tuning prior \cite{Allanach:2006jc, Balazs:2012qc}, without directly
calculating the Jacobian factor.

Nonetheless the conventional tuning remains a very simple and useful measure and has continued to be used widely with the literature. We will employ it here for the following reasons:
\begin{enumerate}
\item It is the most widely used tuning measure with which one can compare;
\item It gives a good approximation of the effective fine tuning prior;
\item It is simple to understand and apply;
\item It provides a better match to the more complicated multi-parameter measure \cite{Athron:2007ry} than combining sensitivities in quadrature.
\end{enumerate}

In particular please note that the simplicity and wide use is very important since this is the first quantitative investigation into tuning in this
model and therefore comparison to what has been done in other models is of greater significance. Applying this measure provides a quantification of the severity of tuning in the model, shows which regions have the least fine tuning and could be used as an “effective fine tuning prior” in future Bayesian studies of the model.


\subsection{Master Formula}

Having concluded the discussion on the motivation and suitability of
this measure we now proceed to apply it in a quantitative analysis of
fine tuning. To do so we first derive and present the master formula
which gives the explicit expression from which the fine tuning is
calculated.  Using Equations~\ref{mz},~\ref{sin2b},~\ref{ms} and~\ref{delta}, we derive this master formula for fine tuning in the
E$_6$SSM\footnote{Note we have left two terms in the second line of Eq.~\ref{Eq:master-formula} written in terms of derivatives of $\cos^2 \beta$ and $\sin^2 \beta$ with respect to $a$.  Substituting for soft masses here would unnecessarily clutter the expression and we note that these terms are numerically negligible since their contribution to fine tuning is very small ($ < \mathcal{O}(1)$). This is due to the fact that they will be multiplied by an overall factor of order $\mathcal{O}(< 10^{-12}).$},

\begin{equation}
\begin{split}
\Delta_a \approx & c^{-1} \times \frac{a }{M_Z^2 (\tan^2\beta - 1)} \bigg\lbrace \frac{ (1- \tan^2\beta)}{2} \frac{ \partial (\lambda^2 s^2) }{ \partial a  } + \frac{ \partial  m_d^2}{ \partial a  }  - \tan^2\beta \frac{\partial m_u^2}{\partial a}  \\
& + \frac{ g_1'^2 }{2} (Q_1 - \tan^2 \beta Q_2) \left( Q_s \frac{ \partial s^2 }{ \partial a} + \frac{4 M_Z^2}{\bar{g}^2} \frac{\partial}{\partial a} (Q_1 \cos^2 \beta + Q_2 \sin^2 \beta) \right) \\
& - \frac{\tan\beta}{\cos 2\beta} \left[1 + \frac{M_Z^2 }{m_d^2 + m_u^2 + \lambda^2 s^2 + \frac{ g_1'^2 }{2} Q_s s^2 ( Q_1 + Q_2 ) } \right] \times  \\
& \times \left[\sqrt{2} \frac{\partial (\lambda A_{\lambda} s) }{\partial a} - \sin 2\beta \frac{\partial}{\partial a} ( m_d^2 + m_u^2 + \lambda^2 s^2 +  \frac{g_1'^2}{2} Q_s ( Q_1 + Q_2 ) s^2 ) \right] \bigg\rbrace,
\end{split}
\label{Eq:master-formula}
\end{equation}
where
\beq c = \left[ 1 - \frac{4}{(\tan^2\beta - 1)}  \frac{g_1'^2 }{\bar{g}^2}(Q_1 - \tan^2 \beta Q_2)\times (Q_1 \cos^2 \beta + Q_2 \sin^2\beta) \right], \eeq
and $\bar{g}^2 =  ({g^{\prime}}^2 + g_2^2). $ For $\tan \beta = 10$; $c^{-1} \simeq 0.88. $

The aim is to expand the low energy parameters, including $s$, in terms of the GUT-scale 
universal input parameters using the E$_6$SSM RGEs as mentioned in the previous section. 
Next, the formula is implemented into a private cE$_6$SSM spectrum generator (described in \cite{Athron:2009bs, Athron:2009ue}) and fine tuning at 
each point in the scanned parameter space is calculated. In order to ensure accuracy of the results, the derivatives 
in the master formula for $a =\lambda(0)$ and $a=\kappa(0)$ are calculated numerically. And in order to calculate, 
\begin{equation}
 \frac{\partial}{\partial a } s^2,
\end{equation}
we use
\begin{equation}
 s^2 = - \frac{2}{g_1'^2 Q_s^2} m_s^2,
\end{equation}
where, as usual, $m_s^2$ is expanded in terms of the GUT parameters.

Finally, throughout our study, we fix $\tan\beta = 10$ since larger and smaller values restrict the availability of $m_h \sim 125$ GeV, 
and the parameter space \cite{Athron:2012sq}. 

\section{Results and discussion}

The scans are taken for fixed $s=5-10$ TeV corresponding to $M_{Z'}=1.9-3.8$ TeV. 
We scan over \beq -3 \lesssim \lambda_3 (0) \lesssim 0
\ \ \text{and} \ \ 0 \lesssim \kappa_1(0) = \kappa_2(0) = \kappa_3(0) \lesssim 3 \eeq while fixing
$\lambda_{1,2} (0) = 0.1$ and $\tan\beta = 10$. The sign of $\lambda \equiv \lambda_3(0)$ is
a free parameter in our convention since we are setting $s$ and $m_{1/2} > 0$. However as with previous studies \cite{Athron:2012sq} we found that most of the parameter space is covered with $\lambda < 0$, while $\lambda > 0$ covers a much smaller region of the parameter space. Therefore we focused on $\lambda < 0$ in our study. The other GUT
parameters: $m_0, m_{1/2}$ and $A_0$ are obtained as an output so that
the EWSB conditions are satisfied to one-loop order.  Then we plot
both $m_h$ and $\Delta_{max}$ in the $m_0-m_{1/2}$ plane. The key at the top-left of all plots corresponding to $m_h$ shows the central value in a bin of width $\pm 0.5$ GeV, while that corresponding to $\Delta$ shows the central value in a bin of width $\pm 50$.

\begin{figure}[H] 
\begin{tabular}{cc}
\resizebox{!}{6.00cm}
{\includegraphics{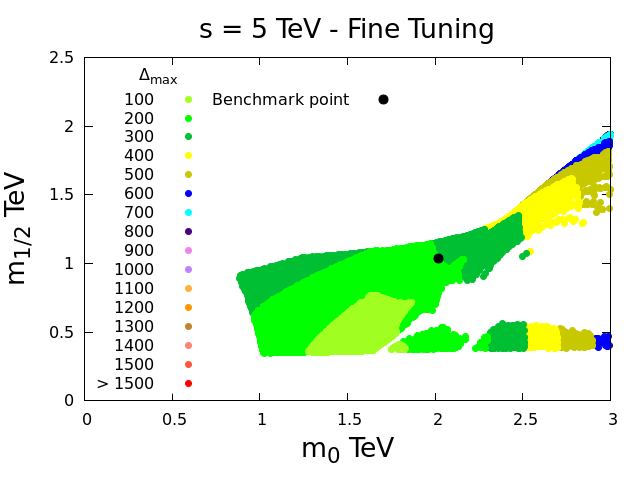}}
\resizebox{!}{6.00cm}
{\includegraphics{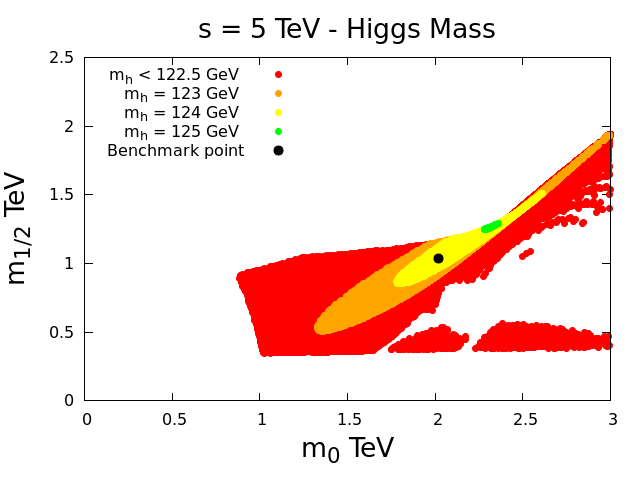}}
\end{tabular}
\caption{ $\Delta_{max}$ (left) and $m_h$ (right) in the $m_0-m_{1/2}$
  plane for $\tan\beta = 10$ and $s=5$ TeV corresponding to
  $M_{Z'}=1.9$ TeV. We also fixed $\lambda_{1,2} (0) =
    0.1$ while scanning over $-3 \leq \lambda_3 (0) \leq 0
    \ \ \text{and} \ \ 0 \leq \kappa_{1,2,3} (0) \leq 3$. The
  benchmark point corresponds to $m_0=2020 , m_{1/2}=1033$ GeV.}
\label{Fig:s5}
\end{figure}

\begin{figure}[h!] 
\begin{center}
\begin{tabular}{cc}
\resizebox{!}{6.00cm}
{\includegraphics{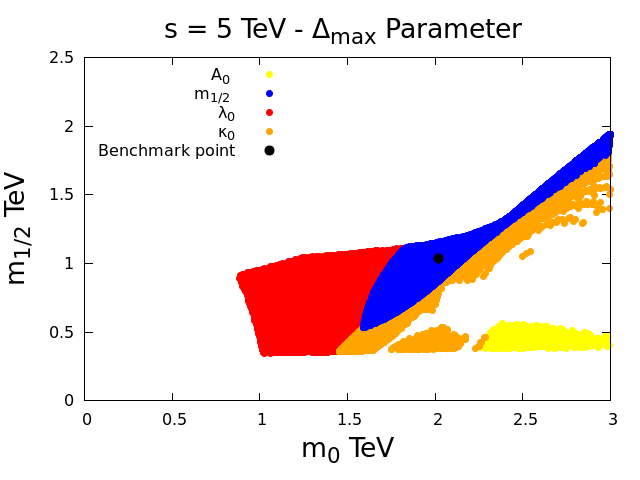}}
\resizebox{!}{6.00cm}
{\includegraphics{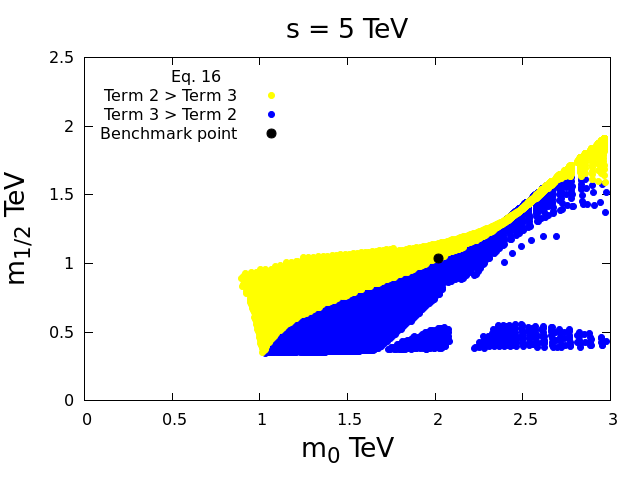}}
\end{tabular}
\caption{The left panel highlights the parameter responsible for
  the largest amount of fine tuning, $\Delta_{max}$, in the
  $m_0-m_{1/2}$ plane for $\tan\beta = 10$ and  $s=5$ TeV corresponding to 
$M_{Z'}=1.9$ TeV.  On the right a coarse scan shows which terms
  Eq.~\ref{mz2} give the largest contribution, with regions where the
  largest contribution comes from term 2, which is proportional to
  $m_d^2 - m_u^2 \tan^2 \beta$, are shown in yellow and while regions
  where the dominant contribution is from term 3, proportional to
  $M_{Z^\prime}^2$ are shown in blue.}
\label{Fig:s5T2T3}
\end{center}
\end{figure}

\begin{figure}[H]
\begin{tabular}{cc}
\resizebox{!}{6.00cm}
{\includegraphics{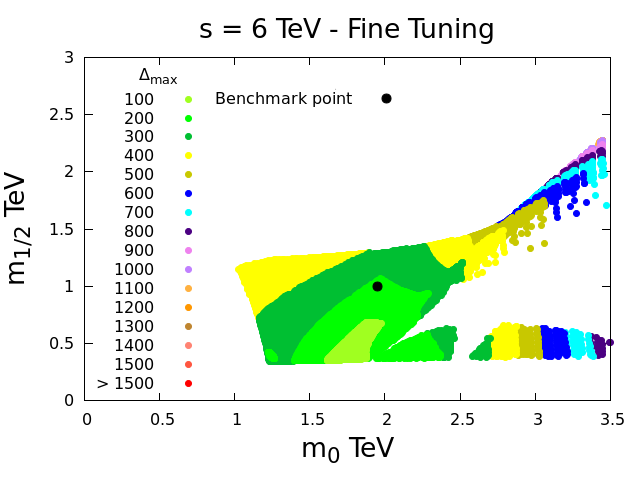}}
\resizebox{!}{6.00cm}
{\includegraphics{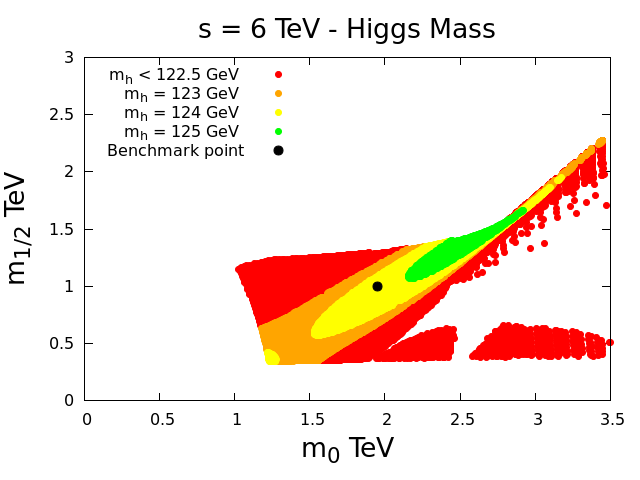}}
\end{tabular}
\caption{$\Delta_{max}$ (left) and $m_h$ (right) in the $m_0-m_{1/2}$ plane
for $\tan\beta = 10$ and  $s=6$ TeV corresponding to 
$M_{Z'}=2.3$ TeV.  The benchmark point corresponds to $m_0=1951 , m_{1/2}=1003$ GeV.}
\label{Fig:s6}
\end{figure}

\begin{figure}[h!] 
\begin{center}
\begin{tabular}{cc}
\resizebox{!}{6.00cm}
{\includegraphics{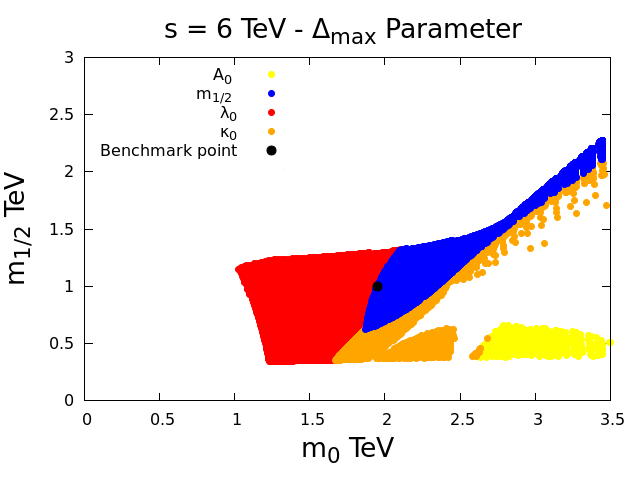}}
\resizebox{!}{6.00cm}
{\includegraphics{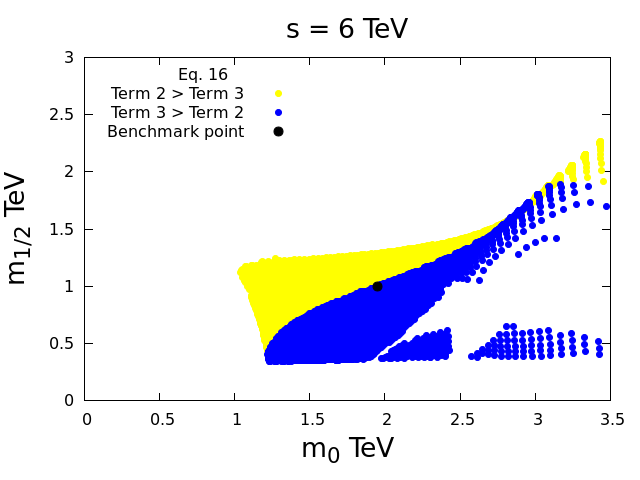}}
\end{tabular}
\caption{The left panel highlights the parameter responsible for
  the largest amount of fine tuning, $\Delta_{max}$, in the
  $m_0-m_{1/2}$ plane for $\tan\beta = 10$ and  $s=6$ TeV corresponding to 
$M_{Z'}=2.3$ TeV.  On the right a coarse scan shows which terms
  Eq.~\ref{mz2} give the largest contribution, with regions where the
  largest contribution comes from term 2, which is proportional to
  $m_d^2 - m_u^2 \tan^2 \beta$, are shown in yellow and while regions
  where the dominant contribution is from term 3, proportional to
  $M_{Z^\prime}^2$ are shown in blue.}
\label{Fig:s6T2T3}
\end{center}
\end{figure}

\begin{figure}[H]
\begin{tabular}{cc}
\resizebox{!}{6.00cm}
{\includegraphics{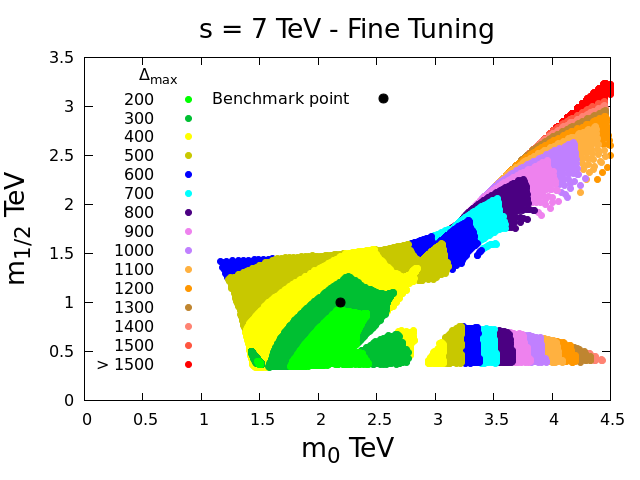}}
\resizebox{!}{6.00cm}
{\includegraphics{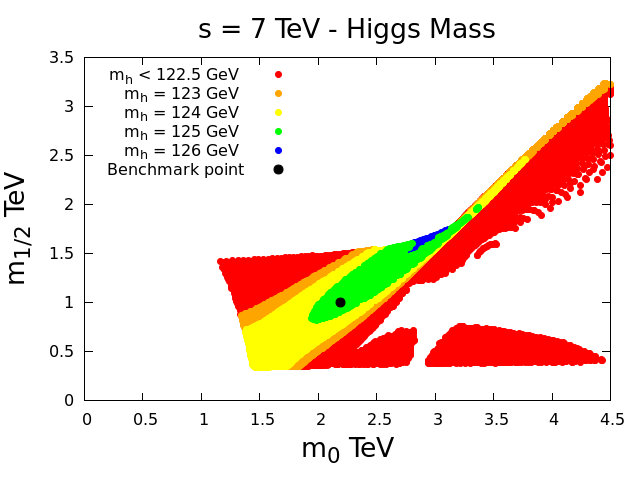}}
\end{tabular}
\caption{$\Delta_{max}$ (left) and $m_h$ (right) in the $m_0-m_{1/2}$ plane
for $\tan\beta = 10$ and  $s=7$ TeV corresponding to 
$M_{Z'}=2.6$ TeV. 
The benchmark point corresponds to $m_0=2186 , m_{1/2}=1004$ GeV.}
\label{Fig:s7}
\end{figure}

\begin{figure}[h!] 
\begin{center}
\begin{tabular}{cc}
\resizebox{!}{6.00cm}
{\includegraphics{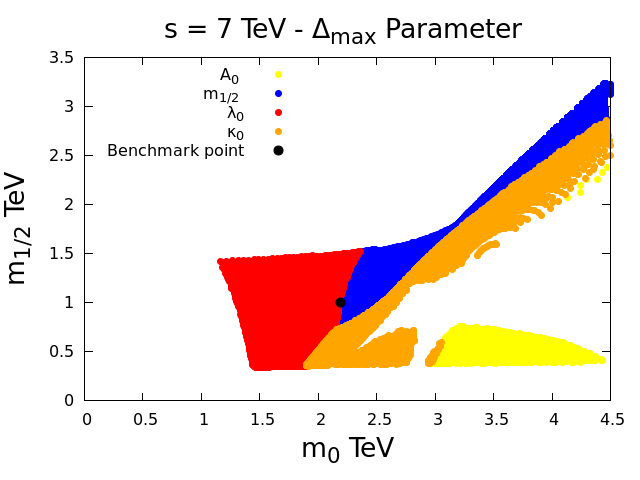}}
\resizebox{!}{6.00cm}
{\includegraphics{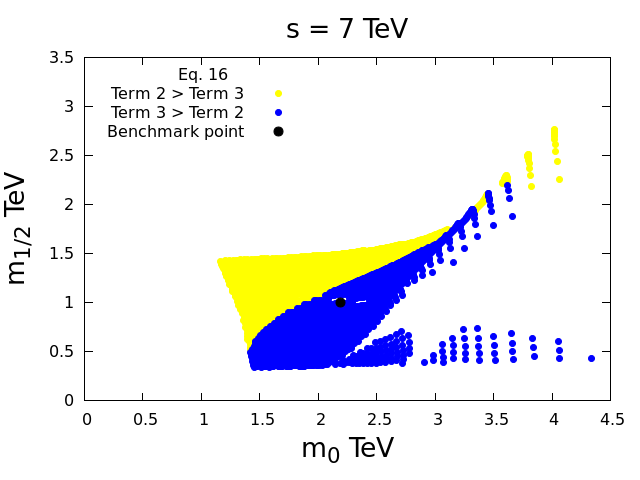}}
\end{tabular}
\caption{The left panel highlights the parameter responsible for
  the largest amount of fine tuning, $\Delta_{max}$, in the
  $m_0-m_{1/2}$ plane for $\tan\beta = 10$ and  $s=7$ TeV corresponding to 
$M_{Z'}=2.6$ TeV.  On the right a coarse scan shows which terms
  Eq.~\ref{mz2} give the largest contribution, with regions where the
  largest contribution comes from term 2, which is proportional to
  $m_d^2 - m_u^2 \tan^2 \beta$, are shown in yellow and while regions
  where the dominant contribution is from term 3, proportional to
  $M_{Z^\prime}^2$ are shown in blue.}
\label{Fig:s7T2T3}
\end{center}
\end{figure}

\begin{figure}[H]
\begin{tabular}{cc}
\resizebox{!}{6.00cm}
{\includegraphics{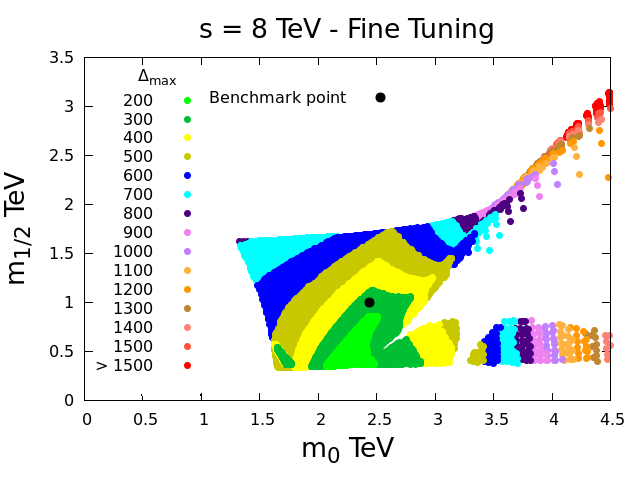}}
\resizebox{!}{6.00cm}
{\includegraphics{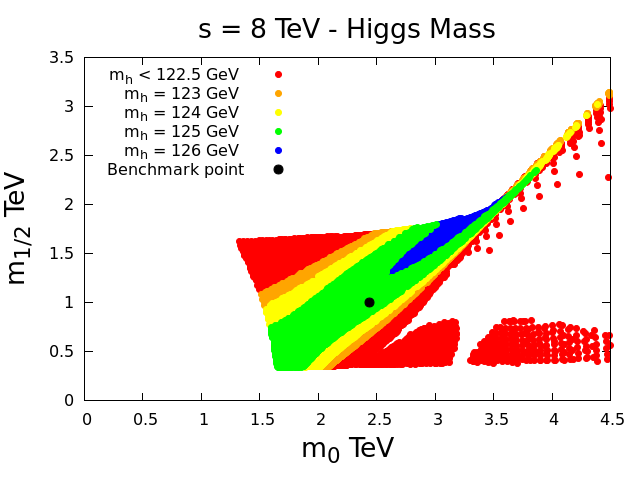}}
\end{tabular}
\caption{$\Delta_{max}$ (left) and $m_h$ (right) in the $m_0-m_{1/2}$ plane
for $\tan\beta = 10$ and  $s=8$ TeV corresponding to 
$M_{Z'}=3.0$ TeV. 
The benchmark point corresponds to $m_0=2441 , m_{1/2}=1002$ GeV.}
\label{Fig:s8}
\end{figure}

\begin{figure}[h!] 
\begin{center}
\begin{tabular}{cc}
\resizebox{!}{6.00cm}
{\includegraphics{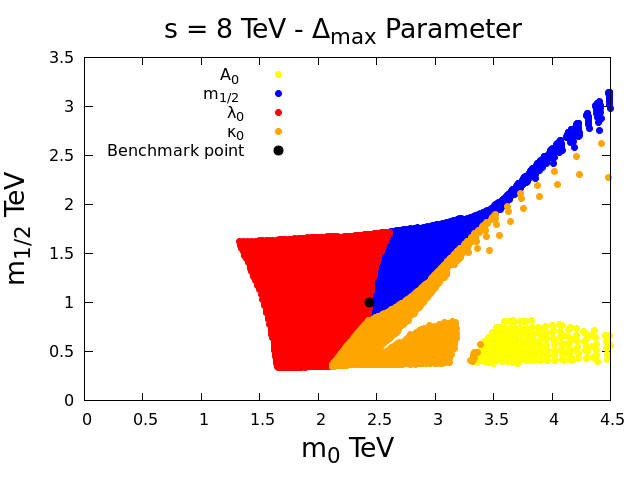}}
\resizebox{!}{6.00cm}
{\includegraphics{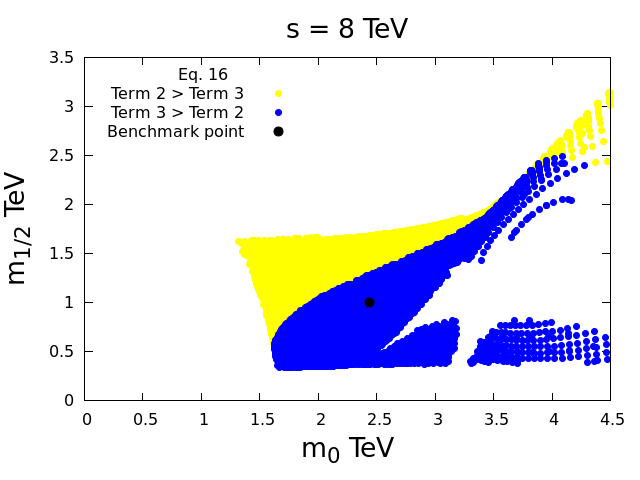}}
\end{tabular}
\caption{The left panel highlights the parameter responsible for
  the largest amount of fine tuning, $\Delta_{max}$, in the
  $m_0-m_{1/2}$ plane for $\tan\beta = 10$ and  $s=8$ TeV corresponding to 
$M_{Z'}=3.0$ TeV.  On the right a coarse scan shows which terms
  Eq.~\ref{mz2} give the largest contribution, with regions where the
  largest contribution comes from term 2, which is proportional to
  $m_d^2 - m_u^2 \tan^2 \beta$, are shown in yellow and while regions
  where the dominant contribution is from term 3, proportional to
  $M_{Z^\prime}^2$ are shown in blue.}
\label{Fig:s8T2T3}
\end{center}
\end{figure}

\begin{figure}[H]
\begin{tabular}{cc}
\resizebox{!}{6.00cm}
{\includegraphics{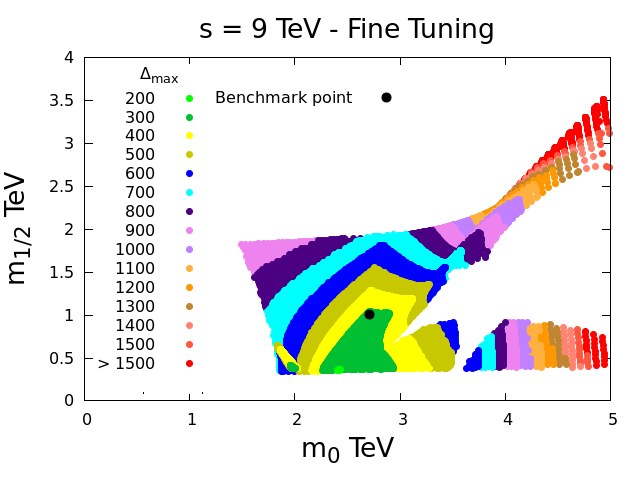}}
\resizebox{!}{6.00cm}
{\includegraphics{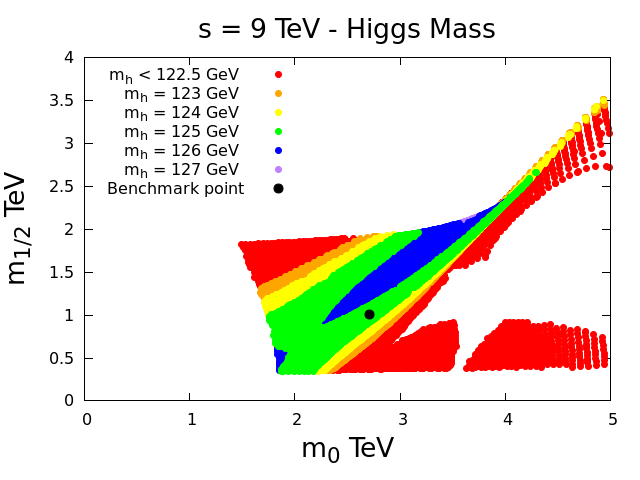}}
\end{tabular}
\caption{$\Delta_{max}$ (left) and $m_h$ (right) in the $m_0-m_{1/2}$ plane
for $\tan\beta = 10$ and  $s=9$ TeV corresponding to 
$M_{Z'}=3.4$ TeV. 
  The benchmark point corresponds to $m_0=2709 , m_{1/2}=1001$ GeV.}
\label{Fig:s9}
\end{figure}

\begin{figure}[h!] 
\begin{center}
\begin{tabular}{cc}
\resizebox{!}{6.00cm}
{\includegraphics{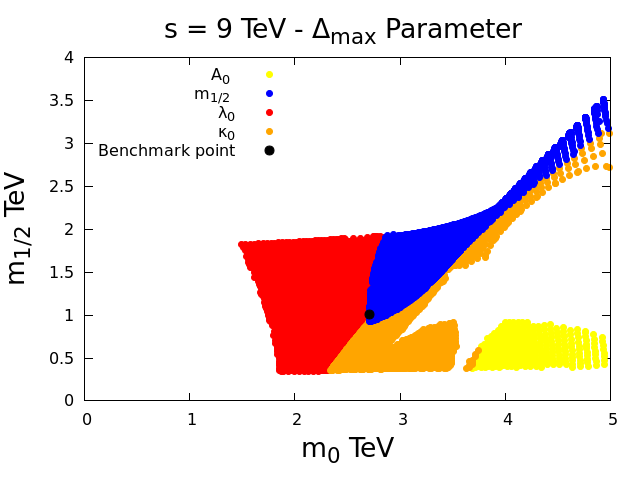}}
\resizebox{!}{6.00cm}
{\includegraphics{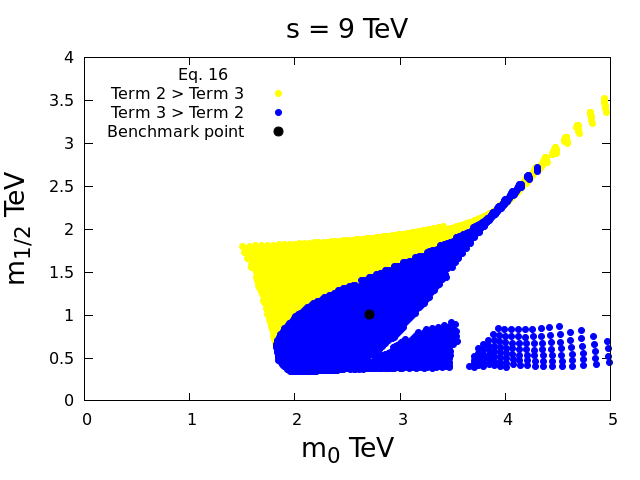}}
\end{tabular}
\caption{The left panel highlights the parameter responsible for
  the largest amount of fine tuning, $\Delta_{max}$, in the
  $m_0-m_{1/2}$ plane for $\tan\beta = 10$ and  $s=9$ TeV corresponding to 
$M_{Z'}=3.4$ TeV.  On the right a coarse scan shows which terms
  Eq.~\ref{mz2} give the largest contribution, with regions where the
  largest contribution comes from term 2, which is proportional to
  $m_d^2 - m_u^2 \tan^2 \beta$, are shown in yellow and while regions
  where the dominant contribution is from term 3, proportional to
  $M_{Z^\prime}^2$ are shown in blue.}
\label{Fig:s9T2T3}
\end{center}
\end{figure}

\begin{figure}[H]
\begin{tabular}{cc}
\resizebox{!}{6.00cm}
{\includegraphics{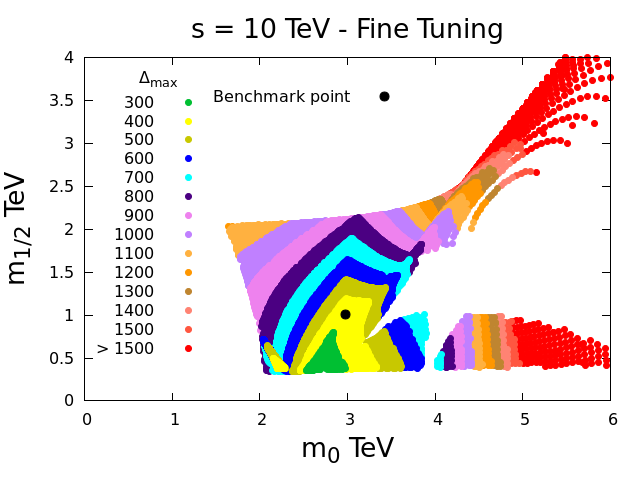}}
\resizebox{!}{6.00cm}
{\includegraphics{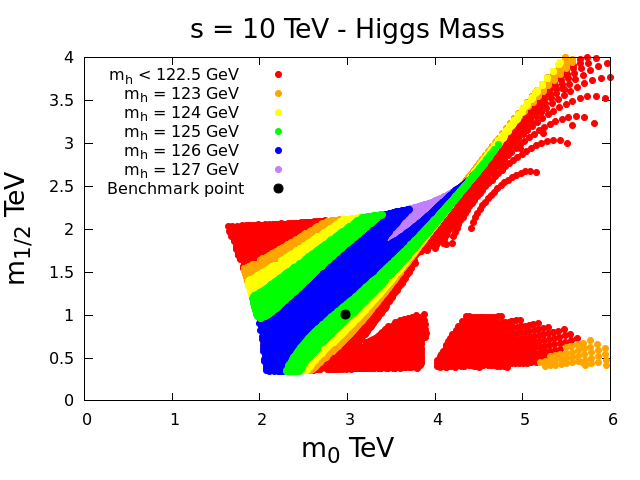}}
\end{tabular}
\caption{$\Delta_{max}$ (left) and $m_h$ (right) in the $m_0-m_{1/2}$ plane
for $\tan\beta = 10$ and  $s=10$ TeV corresponding to 
$M_{Z'}=3.8$ TeV. 
 The benchmark point corresponds to $m_0=2975 , m_{1/2}=1005$ GeV.}
\label{Fig:s10}
\end{figure}

\begin{figure}[h!] 
\begin{center}
\begin{tabular}{cc}
\resizebox{!}{6.00cm}
{\includegraphics{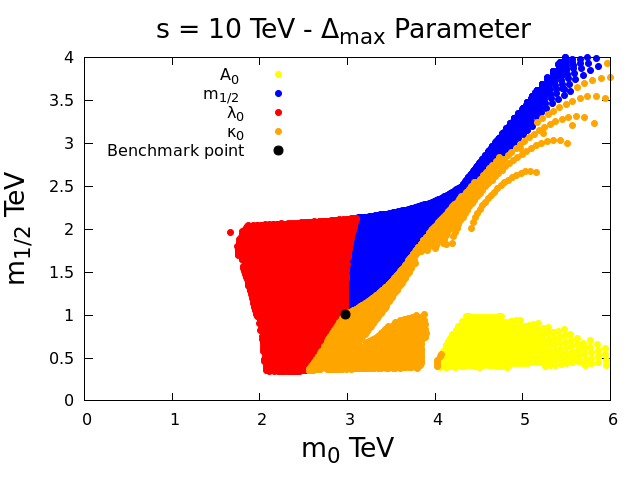}}
\resizebox{!}{6.00cm}
{\includegraphics{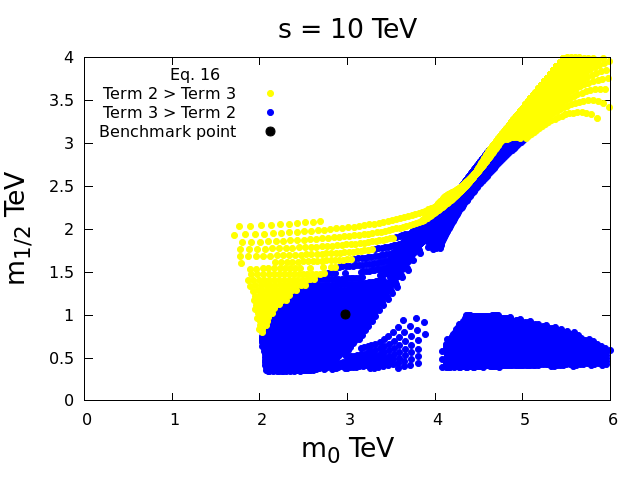}}
\end{tabular}
\caption{The left panel highlights the parameter responsible for
  the largest amount of fine tuning, $\Delta_{max}$, in the
  $m_0-m_{1/2}$ plane for $\tan\beta = 10$ and  $s=10$ TeV corresponding to 
$M_{Z'}=3.8$ TeV.  On the right a coarse scan shows which terms
  Eq.~\ref{mz2} give the largest contribution, with regions where the
  largest contribution comes from term 2, which is proportional to
  $m_d^2 - m_u^2 \tan^2 \beta$, are shown in yellow and while regions
  where the dominant contribution is from term 3, proportional to
  $M_{Z^\prime}^2$ are shown in blue.}
\label{Fig:s10T2T3}
\end{center}
\end{figure}

Moreover, we select a benchmark point corresponding to each value of
$s.$ These points possess the smallest fine tuning in the
$m_0-m_{1/2}$ plane consistent with a Higgs mass within the
$124<m_h<127$ GeV range, and $m_{\tilde{g}} \geq 850$ GeV.  They are denoted as a 
black dot in Figures ~\ref{Fig:s5}-~\ref{Fig:s10T2T3}.  These
points and the relevant physical masses are summarised in Table
~\ref{table:benchmarks} in Appendix~\ref{A}.

In the left panel of Fig.~\ref{Fig:s5} the results for $s=5$ TeV,
corresponding to $M_{Z'}=1.9$ TeV, are shown with fine
tuning contours, ranging from $100$ to above $800$ for the highest
$m_0$. For each value of $m_0$ and
$m_{1/2}$, the parameters $\lambda$, $\kappa$, and $A$ take different values. 
Since the Higgs mass strongly depends both on stop corrections and
$\lambda$, it will also take different values denoted by the Higgs mass contours
displayed in the right panel of Fig.~\ref{Fig:s5}. Since both fine tuning
and the Higgs mass vary over the $m_0-m_{1/2}$ plane the mass of the
Higgs discovered at the LHC plays a crucial rule in fixing the level
of tuning, though this dependence is significantly more complicated
than in the MSSM. Thus, although for $s=5$ TeV the tuning can in principle be
as low as 100, in order to obtain $m_h \sim 124$ GeV the fine tuning must be more than
twice as large as this. A benchmark representing points with the
lowest tuning compatible with data shown as black dot in
Fig.~\ref{Fig:s5} having  $\Delta_{BM} = 251$ with $m_h \approx 124$ GeV. Note that $m_h \sim 125$ GeV is 
almost impossible to achieve for  $s=5$ TeV (represented by the very small green region
in the right panel). In addition, the value $M_{Z'}=1.9$ TeV slightly violates the CMS limit
$M_{Z'}\gtrsim 2.08$ TeV \cite{Chatrchyan:2012it}, although this limit does not take into account the
presence of lighter singlet states which increase the $Z'$ width and reduce the leptonic
branching ratio, weakening this limit as discussed in  \cite{Athron:2011wu}.

One also needs to take into account LHC constraints from squark and gluino
searches which rule out $m_{1/2} \lesssim 1$ TeV 
corresponding to a gluino mass $m_{\tilde{g}}  \lesssim 850$ GeV \cite{Athron:2012sq}. 

In Appendix~\ref{A} we provide a set on benchmark points corresponding to
$m_{1/2} \sim 1$ TeV and these benchmark points are denoted by small black dots on the Figures.
We emphasise that the cE$_6$SSM has not been studied by any of the LHC experiments, and that the gluino mass
limits in the E$_6$SSM may differ from those of the MSSM as discussed recently \cite{Belyaev:2012si}.
Therefore, in choosing our minimum tuning benchmarks, the limits we assumed are quite conservative.
From the results in \cite{Athron:2012sq}, we find that
in the cE$_6$SSM the gluino mass is approximately
given by $m_{\tilde{g}} \sim 0.85 m_{1/2}$ and the first and second generation squark masses are given by
$m_{\tilde{q}}\sim (1.3-1.8)m_0$, depending on $m_{1/2}$. In the future (for example when the full 8 TeV data set is analysed) the allowed values of 
$m_0$ and $m_{1/2}$ are expected to increase according to these approximate relations.  Therefore, we show in Appendix~\ref{B} (Table~\ref{table:mGluinoFT}) the minimum allowed fine tuning associated with gluino mass in the $1 \leq m_{\tilde{g}} \leq 1.5$ TeV range, and the usual range for the singlet VEV $s = 5 - 10$ TeV. Clearly, the fine tuning in the cE$_6$SSM is not as large as that in the CMSSM, where increasing $m_{\tilde{g}}$ to 1.5 TeV leads to minimum fine tuning $> 1000$ as found in ~\cite{Ghilencea:2012gz}, while it varies between $\sim 600 - 800$ in the cE$_6$SSM.

At first sight, the distribution of fine tuning in the $m_0-m_{1/2}$
plane could seem counter intuitive since one might expect the region of
smaller values of $m_0$ and $m_{1/2}$ to possess lower fine tuning.
However, the variation of $\Delta_{max}$ can be understood by studying
which parameter contributes the maximum fine tuning at each point in
the parameter space. We show this in Fig.~\ref{Fig:s5T2T3}
(left panel) where it is clear that the region of small $m_0$ and
$m_{1/2}$ is dominated by large fine tuning in the parameter $\lambda_0$, 
resulting from a large $|\mu_{\text{eff}}|$ term in this region. 

In addition, $\kappa_0$ can contribute to
$\Delta_{{max}}$ since $A_{\lambda}$ and $m_s$ are
strongly dependent on this parameter. 
The physical origin of the fine tuning in $\kappa_0$ is due to the loops of exotic D-particles
which serve to radiatively drive the singlet mass squared negative which triggers electroweak symmetry breaking. Finally,
$m_0$ can be the source of fine tuning for very large values of $m_0$ which is the region extending beyond what we show in the plots.


The relative fine tuning in the input parameters $\{m_0,m_{1/2},A,\lambda(0),\kappa(0)\}$ does not directly tell us any information about the relative importance of the second and third terms on the right-hand side
of Eq.~\ref{mz2}, both of which can independently be large and hence lead to a large $|\mu_{\text{eff}}|$
which is manifested as large fine tuning in $\lambda_0$.
It is therefore instructive to directly compare the magnitudes of the second and third terms 
of Eq.~\ref{mz2}, where the former is proportional to 
$m_u^2$ and $m_d^2$, hence sfermions, and the latter is proportional to $M_{Z^\prime}^2$. 
In Fig.~\ref{Fig:s5T2T3} (right panel) we scan the parameter space for $s=5$ TeV, and for each point we show which 
of the two terms is larger. The larger of the two
would be responsible for the fine tuning at the corresponding point. 
It is clear, then, that $M_{Z^\prime}$ 
(blue region) not only controls the minimum fine tuning allowed, but also is 
the dominating source of fine tuning over large regions of 
the parameter space. This is true for all the other values of $s$. However, some substantial 
contribution to fine tuning comes from sfermions as seen in the yellow region. 

As we increase $s$ to 6 TeV (shown in Fig.~\ref{Fig:s6}), 
we simultaneously satisfy the CMS mass limit on the $Z'$ mass, with $M_{Z'}=2.3$ TeV, 
and we obtain more points with the heavier Higgs mass $m_h=125$ GeV. 
Interestingly, the benchmark point in this case has a fine tuning
$\Delta_{BM} = 233$ for $m_h \approx 124$ GeV which is
slightly smaller than for the previous case with $s=5$ TeV.
Additionally, in the left panel in Fig.~\ref{Fig:s6} a tiny region of $\Delta_{max} = 200$ appears as a small circle inside the $\Delta_{max} = 300$ band. While it is still $\lambda_0$ that is responsible for $\Delta_{max}$ in that area as seen in the left panel in Fig.~\ref{Fig:s6T2T3}, this region is associated with a slightly smaller $|\mu_{\text{eff}}|$ ($|\lambda_0|$) and larger $\kappa_0$ than in the adjacent regions, an effect which was not present in the results of $s=5$ TeV.

 Moreover, Fig.~\ref{Fig:s6T2T3} shows that the origin of fine tuning depends on the 
 point in the $m_0-m_{1/2}$ plane consistent with the Higgs mass and the LHC limits
 of squark and gluino masses, estimated above as $m_{\tilde{g}} \sim 0.85 m_{1/2}$ and 
 $m_{\tilde{q}}\sim (1.3-1.8)m_0$. For example if the squark and gluino masses are 
 increased then it is possible that fine tuning is dominated by fine tuning in $m_{1/2}$
 or in $\lambda_0$ via large $|\mu_{\text{eff}}|$ which could be due to heavy stop masses 
 rather than large $M_{Z^\prime}$ according to the right panel in Fig.~\ref{Fig:s6T2T3}.

For $s=7$ TeV, corresponding to $M_{Z'}=2.6$ TeV, the region with
$m_h \sim 125$ GeV expands in comparison to $s=5$ and $6$ TeV,
as can be seen by comparing the right panel in Fig.~\ref{Fig:s7}, to the
previous plots. In addition a very small region with $m_h \sim 126$ GeV appears for the first time.
In the left panel of Fig.~\ref{Fig:s7}, fine
tuning starts from 200, and reaches 600 outside the
middle region. In addition, the tiny circle of points with smaller fine tuning than its surroundings in the small $m_0-m_{1/2}$ region, which appeared previously in the results for $s=6$ TeV, now grows a little.

The chosen benchmark point has $\Delta_{BM} = 270$ for $m_h
\approx 125$ GeV.  Notice how increasing $s$, hence $M_{Z'}$, affects
the lowest fine tuning possible in the parameter space, confirming
that it is the $M_{Z^\prime}$ term in Eq.~\ref{mz2} dominating fine
tuning and defining its lowest value as can be seen in the right panel of Fig.~\ref{Fig:s7T2T3}.
As before, this conclusion depends on the particular point in the $m_0-m_{1/2}$ plane.

For $s=8$ TeV the Higgs mass $m_h \sim 125$ GeV
dominates over most of the $m_0-m_{1/2}$ plane as shown in the right panel of Fig.~\ref{Fig:s8}. 
Also the $m_h \sim 126$ GeV region has become larger.
However, fine tuning starts from 300, and the portion of the parameter space with $\Delta_{max} \geq 500$ 
is now more apparent than in the $s=7$ TeV case. 
The Benchmark point has $\Delta_{BM} = 302$ for $m_h \approx 125$ GeV. 
The dominance of the $M_{Z^\prime}$ term in Eq.~\ref{mz2} for fine
tuning can be seen in the right panel of Fig.~\ref{Fig:s8T2T3}, with 
this conclusion dependent on the particular point in the $m_0-m_{1/2}$ plane.

As we reach $s=9$ TeV, corresponding to $M_{Z'}=3.4$ TeV, which is
shown in Fig.~\ref{Fig:s9}, we see that the region where $m_h \sim
125$ GeV starts to shrink and is replaced by $m_h \sim 126 $ GeV. If
the Higgs mass is indeed $m_h \sim 126 $ GeV then there is a
preference for $s=9$ TeV,
especially for smaller values of $m_0$ and $m_{1/2}$.  This
illustrates the importance of an accurate determination in the Higgs
mass for selecting the most appropriate value of $s$.  Fine tuning
starts from 200, although a very small region, and quickly increases to 500 such that a significant
portion of the parameter has $\Delta_{max} \gtrsim 500$. The benchmark
point has $\Delta_{BM} = 330$ for $m_h \approx 125$ GeV.  The
dominance of the $M_{Z^\prime}$ term in Eq.~\ref{mz2} for fine tuning
can be seen in the right panel of Fig.~\ref{Fig:s9T2T3}, as usual
dependent on the particular point in the $m_0-m_{1/2}$ plane.

Finally, for $s=10$ TeV, corresponding to $M_{Z'}=3.4$ TeV, in the left panel of Fig.~\ref{Fig:s10}
the fine tuning starts from 300, and
the parameter space is severely restricted in terms of fine tuning as
it is mostly covered by points with $\Delta_{max} > 500$. In
addition, the region of $m_h \sim 125$ GeV has shrunk and now occupies a smaller
portion than the $m_h \sim 126$ GeV region.  
In addition a small region with  $m_h \sim 127$ GeV now exists prominently
for the first time (only a miniscule region existed for $s=9$ TeV). 
Moreover, as seen before, the left panel in Fig.~\ref{Fig:s10}
contains short lines of points in the small $m_0-m_{1/2}$ region with smaller fine tuning than their surrounding points for the same reason as before, namely that $|\mu_{\text{eff}}|$ can be somewhat smaller.

The benchmark point has fine tuning
$\Delta_{BM} = 359$ and $m_h \approx 125$ GeV.
The dominance of the $M_{Z^\prime}$ term in Eq.~\ref{mz2} for fine
tuning can be seen in the right panel of Fig.~\ref{Fig:s10T2T3}, with the familiar 
dependence on the particular point in the $m_0-m_{1/2}$ plane.

\section{Conclusion}
Supersymmetric unified models in which the singlet VEV is responsible simultaneously both for $ \mu_{\text{eff}}$
and for the $Z^\prime$ mass,
as in the $E_6$ class of models for example, 
have relatively large fine tuning which is typically dominated by the experimental mass limit on the $Z^\prime$.
To illustrate this, we have investigated the degree of fine tuning throughout the parameter
space of the cE$_6$SSM.
In fact this is the first time
that fine tuning has been studied in any $E_6$ model containing a TeV scale $Z^\prime$.

To quantify fine tuning we have derived a fine tuning master formula for the E$_6$SSM 
and implemented it in a spectrum generator for the constrained
version of the model. Using this we scanned the parameter space of the
cE$_6$SSM.
The results are presented in the $m_0-m_{1/2}$ plane for fixed $\tan \beta =10$ and various $s$
values corresponding to $M_{Z'}\sim 2-4$ TeV.
This value of  $\tan \beta =10$ is the optimum choice for achieving a large enough Higgs mass
in the cE$_6$SSM and so we have exclusively focussed on it here.
We selected benchmark points corresponding to each value of
$s$ which possess the smallest fine tuning while allowing 
a Higgs mass within the $124<m_h<127$ GeV range, and $m_{\tilde{g}} \geq 850$ GeV.  They are
the black dot points in Figures~\ref{Fig:s5}-~\ref{Fig:s10T2T3}.  These benchmark
points and the relevant physical masses are summarised in Table~\ref{table:benchmarks}
for a gluino mass of about $900$ GeV. Table~\ref{table:mGluinoFT} shows how the minimum fine tuning changes as the gluino mass limit increases up to $1.5$ TeV. As remarked earlier, the fine tuning in the cE$_6$SSM is always significantly smaller than that in the cMSSM, for all gluino masses.

It is clear that the $Z^\prime$ mass (determined by the $s$ VEV value) 
has a significant effect on the naturalness of the cE$_6$SSM
model, with higher values leading to increased fine tuning. Therefore future improved direct mass
limits on the $Z^\prime$ mass from the LHC will imply higher fine tuning. 
We have also seen an indirect relation between the Higgs boson mass and
the $Z^\prime$ mass. For example if the Higgs mass turns out to be 
$m_h\gtrsim 127$ GeV
then we are driven to $s\gtrsim 10$ TeV corresponding to $M_{Z'}\gtrsim 3.8$ TeV
requiring higher fine tuning. Conversely if the Higgs mass turns out to be 
$m_h\lesssim 124$ GeV
then $s\gtrsim 5$ TeV corresponding to $M_{Z'}\gtrsim 1.9$ TeV
allowing lower fine tuning. 

Given present limits, the results in Figures~\ref{Fig:s5}-~\ref{Fig:s10T2T3} and Table~\ref{table:benchmarks}
show that the  present lowest value of fine tuning in the cE$_6$SSM,
consistent with a Higgs mass $m_h \sim 125$ GeV, 
varies from $\Delta  \sim 200-400$ where the allowed lowest fine tuning values,
 taking into account the relevant experimental bounds, are dominated by $M_{Z'}$
rather than the other sources of fine tuning.
This is presently significantly lower than the fine tuning in the cMSSM
of $\Delta  \sim 1000$ arising from the large stop masses required to achieve the Higgs mass. 

In the future, the LHC lower limits on gluino and squark masses will improve, along with the $Z^\prime$ mass
limit (or else a discovery will be made) and the Higgs boson mass will be more accurately specified. It is not completely clear where the dominant source of fine tuning in the cE$_6$SSM will originate from in future.
However the results in this paper allow this question to be addressed.
The future $Z^\prime$ mass limit will determine
the minimum $s$ value permitted, while the Higgs mass and gluino and squark mass limits will
determine the allowed regions of the $m_0-m_{1/2}$ plane, from which the fine tuning may be read off
from the contour plots we provide.

\section*{Acknowledgements}
We would like to thank Jonathan P. Hall and Kai Schmidt-Hoberg for helpful discussions during early stages of this study. The work of MB is funded by King Saud University (Riyadh, Saudi Arabia).
SFK acknowledges partial support 
from the STFC Consolidated ST/J000396/1 and EU ITN grants UNILHC 237920 and INVISIBLES 289442 . 
The work of PA is supported by the ARC Centre of Excellence for Particle Physics at the Terascale. PA would also like to thank Tony Williams for reading the manuscript and offering helpful comments.

\appendix
\section{cE$_6$SSM Benchmark points}
\label{A}
Table~\ref{table:benchmarks} lists the details on the masses and parameters
associated with each benchmark (BM) point that was chosen. We can see
that $m_0$ increases significantly as $s$ ($M_{Z'}$) becomes larger,
while $m_{1/2}$ is roughly constant. Upon choosing a BM point, we
imposed the limit $m_{1/2} > 1$ TeV to have gluino mass $m_{\tilde{g}}
> 850$ GeV. The gluino masses for our benchmark points are about 900
GeV or close to it, hence if the experimental limits on
$m_{\tilde{g}}$ are to be increased for constrained models, then fine
tuning will increase as well.  The lightest stop, $\tilde{t}_1$,
masses range from 1.7 TeV to 2.4 TeV for the range of $s$ we studied,
and thereby is above the experimental limits.

\begin{table}[H] 
\begin{center}
\scalebox{0.8}{
 \begin{tabular}{| r | c | c | c | c | c | l | }
  \hline                        
   & BM1 & BM2 & BM3 & BM4 & BM5 & BM6 \\
  \hline 
  $s$ [TeV] & 5 & 6 & 7 & 8 & 9 & 10  \\
 
  $\tan\beta$ & 10 & 10 & 10 & 10 & 10 & 10 \\
    
  $\lambda_3 (M_X)$ & -0.2284 & -0.2646 & -0.25 & -0.2376 & -0.2260 & -0.2171 \\ 
  
  $\lambda_{1,2} (M_X)$ & 0.1 & 0.1 & 0.1 & 0.1 & 0.1 & 0.1 \\

  $\kappa_{1,2,3} (M_X)$ & 0.1760 & 0.1923 & 0.2111 & 0.2288 & 0.2452 & 0.2601 \\

  $m_{1/2}$ [GeV] & 1033 & 1003 & 1004  &  1002 & 1001 &  1005 \\
 
  $m_0$ [GeV] & 2020 & 1951 & 2186 & 2441 & 2709 &  2975 \\

  $A_0$ [GeV] & -83 & 500 & 661 & 781 &  846 & 888 \\
  \hline   

  $m_{\tilde{D_1}}(1,2,3)$ [GeV] & 2252 & 2234 & 2659 & 3149 & 3680 & 4222  \\
  
  $m_{\tilde{D_2}}(1,2,3)$ [GeV] & 3186 & 3501 & 3991 & 4499 & 5017 & 5540  \\
  
  
  
  
  $\mu_D(1,2,3)$ [GeV] & 1782 & 2238 & 2752 & 3279 & 3812 & 4347  \\
  \hline
  
   $|m_{\chi^0_6}|$ [GeV] & 1973 & 2349 & 2727 & 3105 & 3483 & 3861  \\
   
   $m_{h_3} \simeq M_{Z'}$ [GeV] & 1889 & 2267 & 2645 & 3023 &  3401  & 3779  \\

   $|m_{\chi^0_5}|$ [GeV] & 1809 & 2189 & 2566 & 2944 & 3322 & 3699  \\

  \hline
  
  $m_s(1,2)$ [GeV] & 2448 & 2548 & 2897 & 3263 & 3639 & 4014  \\
  $m_{H_2}(1,2)$ [GeV] & 1970 & 1847 & 2023 & 2218 & 2426.5 & 2633  \\
  $m_{H_1}(1,2)$ [GeV] & 1887 & 1685 & 1824 & 1986 & 2167 & 2343  \\
  $\mu_{\tilde{H}}(1,2)$ [GeV] & 492 & 569 & 642 & 711 & 777 & 841  \\
  \hline
  
    $m_{\tilde{u_1}}(1,2)$ [GeV] & 2505  & 2461 & 2687 & 2934 & 3199 & 3468  \\
  $m_{\tilde{u_1}} \simeq m_{\tilde{d_1}}(1,2)$ [GeV] & 2553 & 2507 & 2729 & 2973 & 3235 & 3501  \\
  $m_{\tilde{d_2}}(1,2)$ [GeV] & 2571 & 2558 & 2810 & 3082 & 3372 & 3665  \\
  $m_{\tilde{e_1}}(1,2,3)$ [GeV] & 2136 & 2107 & 2366 & 2641 & 2935 & 3224  \\
  $m_{\tilde{e_2}}(1,2,3)$ [GeV] & 2267 & 2271 & 2550 & 2848 & 3159 & 3468  \\
  $m_{\tilde{\tau_1}}$ [GeV] & 2119 & 2090 & 2347  & 2623 & 2912 & 3200   \\
  $m_{\tilde{\tau_2}}$ [GeV] & 2259 & 2263  & 2541 & 2838 & 3148 & 3457   \\
  $m_{\tilde{b_1}}$ [GeV] & 2202 & 2151 & 2340 & 2549 & 2777 & 3009  \\
  $m_{\tilde{b_2}}$ [GeV] & 2552 & 2539 & 2789 & 3059 & 3347 & 3639  \\
  $m_{\tilde{t_1}}$ [GeV] & 1741 & 1681 & 1839 & 2016 & 2212 & 2411  \\
  $m_{\tilde{t_2}}$ [GeV] & 2215 & 2166 & 2354 & 2561 & 2787 & 3018 \\
  \hline

  $|m_{\chi^0_{3,4}}| \simeq |m_{\chi^\pm_2}| $ [GeV] & 887 & 1174 & 1258 & 1329 & 1386 & 1443  \\
  
  $m_{h_2} \simeq m_A \simeq m_{H^\pm} $ [GeV] & 1890 & 2268 & 2646 & 3025 & 3403 & 3782  \\
  
  $m_h$ [GeV] & 124 & 124 & 125 & 125 & 125 & 125 \\
  \hline

  $m_{\tilde{g}}$ [GeV] & 901 & 879 & 887 & 892 &  898 & 906 \\
  
  $|m_{\chi^\pm_1}| \simeq |m_{\chi^0_2}| $ [GeV] & 285 & 279 & 279 & 279 & 279 & 280  \\
  
  $|m_{\chi^0_1}| $ [GeV] & 162 & 157 & 158 & 158 & 158 & 158  \\
  \hline

  $\Delta_{max}$ & 251 & 233 & 270 & 302 &  330  &  359 \\
  \hline
  
\end{tabular}}
\caption{Parameters and masses for
the benchmarks with lowest fine tuning and Higgs masses in the range of $m_h = 124-125$ GeV in the cE$_6$SSM. 
}
\label{table:benchmarks}
\end{center}
\end{table}

\section{Fine tuning and $m_{\tilde{g}}$}
\label{B}
As the lower limits on the gluino mass are expected to rise, Table~\ref{table:mGluinoFT} shows the minimum amount of the fine tuning corresponding to different values of gluino mass within $m_{\tilde{g}} = 1-1.5$ TeV, and for $s = 5 -10$ TeV. The corresponding Higgs mass is shown in parenthesis next to each value of fine tuning.   

 
\begin{table}[H] 
\begin{center}
\scalebox{0.9}{  
\begin{tabular}{c|c|c|c|c|c|c|}
\cline{1-7}
\multicolumn{1}{ |c  }{$s$ [TeV]}                        &
\multicolumn{1}{ |c| }{5} & 6 & 7 & 8 & 9  & 10   \\ \cline{1-7}
\multicolumn{1}{ |c| }{$m_{\tilde{g}}$ [TeV]} &
\multicolumn{6}{ |c| }{$\Delta$ \ \ \ ($m_h$ [GeV])} \\ \cline{1-7}
\multicolumn{1}{ |c  }{1}                        &
\multicolumn{1}{ |c| }{293 (124)} & 297 (124) & 324 (125) & 367 (125) & 405 (126)  & 443 (126)   \\ \cline{1-7} 
\multicolumn{1}{ |c  }{1.1}                        &
\multicolumn{1}{ |c| }{388 (125)} & 348 (124) & 358 (124) & 408 (125) & 454 (126)  & 497 (126)   \\ \cline{1-7}
\multicolumn{1}{ |c  }{1.2}                        &
\multicolumn{1}{ |c| }{474 (124)} & 440 (125) & 400 (124) & 448 (125) & 500 (126)  & 550 (126)   \\ \cline{1-7}
\multicolumn{1}{ |c  }{1.3}                        &
\multicolumn{1}{ |c| }{ -} & 556 (125) & 462 (124) & 484 (124) & 547 (126)  & 600 (126)   \\ \cline{1-7}
\multicolumn{1}{ |c  }{1.4}                        &
\multicolumn{1}{ |c| }{-} & 658 (125) & 617 (126) & 525 (124) & 587 (125)  & 650 (126)   \\ \cline{1-7}
\multicolumn{1}{ |c  }{1.5}                        &
\multicolumn{1}{ |c| }{-} & - & 767 (125) & 635 (125) & 628 (125)  & 699 (126)   \\ \cline{1-7}

\end{tabular}}
\caption{\footnotesize For different values of the singlet VEV ($s= 5-10$ TeV) corresponding to $M_{Z^{\prime}} \sim 2 - 3.8$ TeV, the effect of rising the lower limit on the gluino mass between $m_{\tilde{g}} = 1-1.5$ TeV on fine tuning is shown. Next to every fine tuning value, the corresponding Higgs mass (in GeV) is shown between parentheses. The dash means there's no $m_h \sim 124 - 127$ GeV found in the scanned parameter space.  }
\label{table:mGluinoFT}
\end{center}
\end{table}

\end{document}